\documentclass[twocolumn,prb,nobibnotes,altaffilletter,amsmath,amssymb,amsfonts]{revtex4}
\usepackage[latin1]{inputenc}
\usepackage{graphicx}
\usepackage{amsmath}
\usepackage{latexsym}
\usepackage{epsfig}
\usepackage{color}
\usepackage[breaklinks=false]{hyperref}

\begin{document}

 \begin{@twocolumnfalse}
 \begin{center}
    \mbox{
{
\color{blue} \href{http://dx.doi.org/10.1039/C3SM51223A}{DOI: 10.1039/C3SM51223A}
}
    }
\end{center}
  \end{@twocolumnfalse}

\title{Motility fractionation of bacteria by centrifugation}

\author{Claudio Maggi$^1$}
\email{claudio.maggi@roma1.infn.it}
\author{Alessia Lepore$^2$}
\author{Jacopo Solari$^1$}
\author{Alessandro Rizzo$^1$}
\author{Roberto Di Leonardo$^1$}

\affiliation{ $^1$Dipartimento di Fisica, Universit\`a di Roma ``Sapienza'', I-00185, Roma, Italy }
\affiliation{ $^2$Dipartimento di Matematica e Fisica, Universita' degli Studi di Roma Tre, Roma, Italy
and CNR-IPCF UOS Roma, Italy}
\date{\today}
\begin{abstract}

Centrifugation is a widespread laboratory technique used to separate mixtures into fractions characterized by a specific size, weight or density. We demonstrate that centrifugation can be also used to separate swimming cells having different motility.  To do this we study self-propelled bacteria under the influence of an external centrifugal field. Using dynamic image correlation spectroscopy we measure the spatially resolved motility of bacteria after centrifugation. A significant gradient in swimming-speeds is observed for increasing centrifugal speeds. Our results can be reproduced by a model that treats bacteria as ``hot" colloidal particles having a diffusion coefficient that depends on the swimming speed.    

\end{abstract}

\maketitle

\section{Introduction} 
Flagellar motility plays a fundamental biological role in prokaryotic and eukaryotic unicellular organisms.
Modulating flagellar activity in response to a variety of chemical and physical stimuli, single-celled micro-organisms can effectively search for optimal environmental conditions~\cite{Phototaxis}. Flagellar motility also plays an important role in medicine, being a major contributing factor to pathogenicity and colonization in bacteria like {\it Vibrio cholerae} \cite{vibrio,Salmonella,Virulence}. 
More recently, the possibility of exploiting swimming micro-organisms as actuators for micro-structures has extended the interest for flagellar motility to the physical domain of  micro-engineering applications~\cite{MagnetoControl,Funnels,ActRat, RatchetSim, Ratchet, Shuttle, ActDep}.
Recognizing the primary role of motility has led to the development of new tools that are capable of a precise and quick characterization of the dynamical properties of cells. Image correlation techniques, as dynamic image correlation spectroscopy (ICS) and differential dyamic microscopy (DDM), are promising tool offering a high-throughput method for characterizing the motility of microorganisms \cite{ICS1,DDM1,DDM2,DDM3,DDM4,DDM5}. 

However, in conjunction to physical tools for motility quantification, it is also desirable to develop physical techniques for sorting colonies, that usually display a high motility variation, into spatially separated fractions characterized by a motility gradient.  Such fractions could be further characterized or isolated for those applications that strongly rely on a highly motile sample. 
Field-flow fractionation has been used to isolate motile nonchemotactic cells from a mixture of different strains \cite{nochem}.
More recently, advances in micro-fabrication techniques have led to the development of microstructures capable of concentrating motile cells~\cite{Funnels, ratchpatt} and sorting them according to their length~\cite{lensort}. 
Fractionation by centrifugation is a widely used technique in biology and chemistry~\cite{Centrifuge}. It relies on the strong sensitivity of sedimentation speed on particle size and composition and therefore allows to separate components according to size, mass or density. For small enough centrifugal accelerations, a densely packed pellet is never attained due to thermal agitation favoring the larger entropy state of a uniform distribution. As a result a stationary state is eventually reached where, each solute is distributed according to the Boltzmann law:

\begin{equation}
\rho(z)\propto \exp\left[-v_d z/D\right]
\label{boltz}
\end{equation} 

\noindent with $v_d=\mu \Delta m a$ is the drift speed induced by a uniform centrifugal acceleration $a$ on a particle having a buoyant mass $\Delta m$, mobility $\mu$ and a diffusion coefficient $D=\mu k_B T$.  Although bacteria will display some variations in the buoyancy mass, we do not expect it to be strongly correlated to their motility. On the other hand swimming bacteria do not rely on thermal agitation for motion, but have their own source of propelling power that makes them a strongly out of equilibrium system. It has been found that in many respects, they can be thought of as ``hot colloids",  with an effective diffusivity that strongly depends on motility and that is typically hundreds of times larger than the Brownian diffusivity of non motile cells \cite{CatesSed, CatesRev}. 
This is also true for active colloidal particles whose sedimentation under gravity as been described by an effective temperature~\cite{JanusGrav}.

Here we demonstrate that a sample of motile {\it E. coli} bacteria, displaying a broad spectrum of swimming speeds, behaves like a mixture of ``hot" colloids having a corresponding broad spectrum of effective temperatures and therefore sedimentation lengths. As a consequence, after centrifugation, non-motile bacteria will accumulate to the bottom of the cell while higher regions are populated with bacteria having an increasing motility/temperature. We used ICS to perform space-resolved motility measurements of bacteria observed over a field of view spanning 1 mm. Space dependent motility distribution were retrieved for centrifugal fields in the range $\sim$ 4-12 $g$ and accounted for with a simple theoretical model of active diffusion.

\section{Experiment} 

\textit{E. coli} cells (MG1655) were grown overnight at
33$^\circ$C in tryptone broth (TB, Difco) containing 1 \% tryptone and
0.5 \% NaCl. The saturated culture was then diluted 1:100 (50 $\mu$l
in 5 ml) into fresh medium and grown at 33$^\circ$C until OD600 = 0.4 
(optical density at 600 nm wavelength) was reached.
Bacterial cells were then harvested from culture media by centrifugation at 2200 rpm for 10 minutes at room temperature. The pellet was resuspended by gently
mixing in a pre-warmed motility buffer composed of 10 mM potassium
phosphate, 0.1 mM Na-EDTA (pH 7.0), 76 mM NaCl and 0.002 \% Tween-20~\cite{prep}. Motility buffer does not sustain cell replication at room temperature, therefore the population remains constant throughout the experiment.

\begin{figure}
\begin{center}
\includegraphics[width=9.cm]{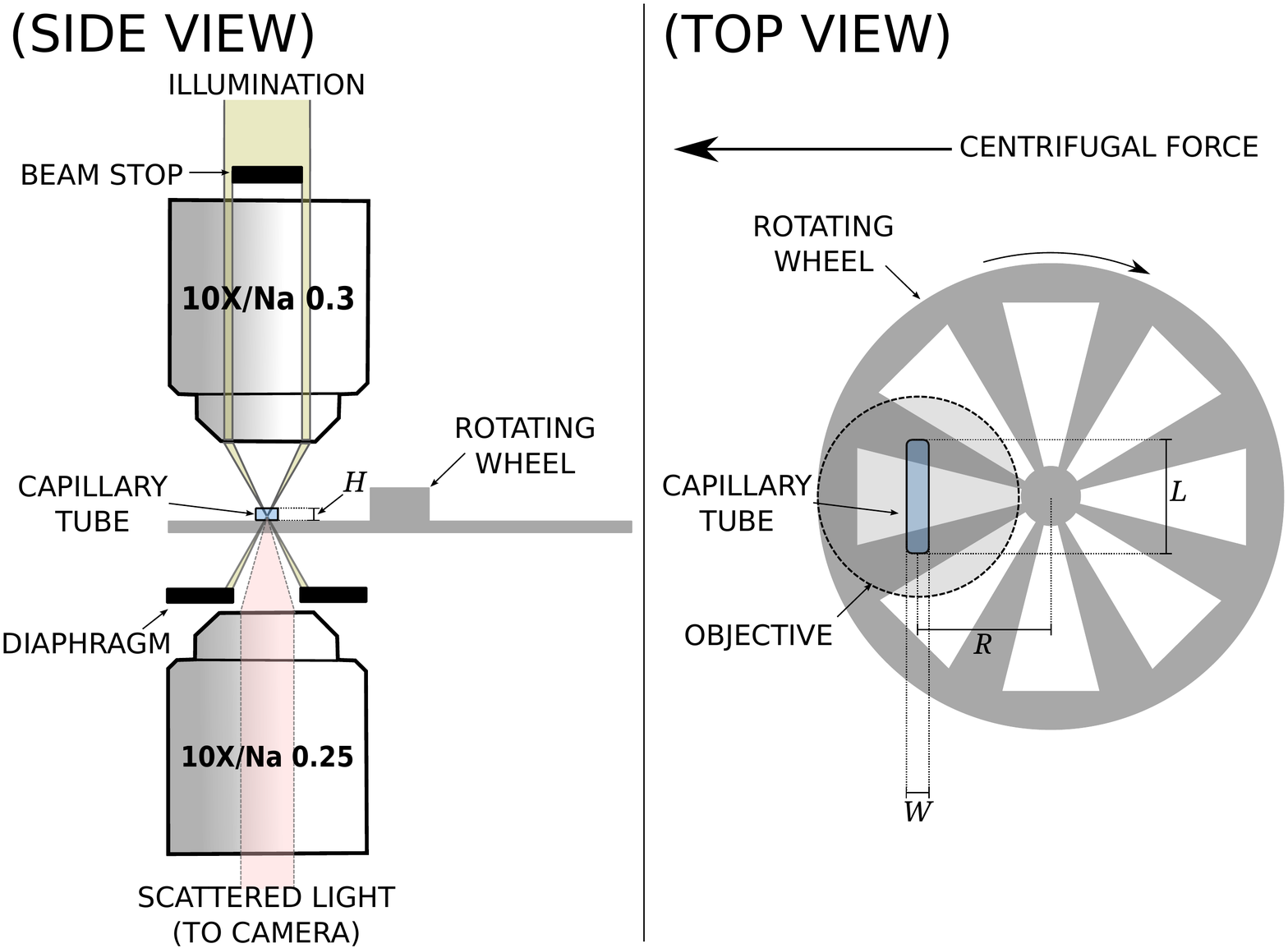}
\caption{ 
Experimental set-up combining dark-field microscopy and a centrifugation apparatus.
(Left) In dark-field configuration a partially blocked illumination beam is focused onto the sample that scatters the incoming radiation. The illumination light is blocked and only the scattered light is collected by the imaging objective.
(Right) The sample is contained in a capillary glass tube attached to a wheel that is driven by a precision motor.}
\label{fig:f0}
\end{center}
\end{figure}

The experimental set-up is designed to combine a dark-field microscope~\cite{MicroBook} with a centrifugation apparatus~\cite{Centrifuge} as shown in Fig.~\ref{fig:f0}.
Large images ($1800 \times 1200$ pixels, i.e. $\sim \, 0.65 \times 0.44 \mathrm{mm}$) are collected at a frame-rate of 33 fps by a CMOS camera (Hamamatsu Orca Flash 2.8) for a total measurement time of about 60 s ($\sim 2000$ images). 
The bacteria are contained in a capillary glass tube (Vitrocom) with near-squared edges with internal dimension $1 \times 0.1 \times 10$ mm ($W \times H \times L$ in Fig.~\ref{fig:f0}). 
Note that our images ($\sim 0.65 \, \mathrm{mm}$ long) cover about half of the capillary width $W$ that is oriented along the centrifugal force field.
The capillary tube is sealed with a biologically inert adhesive wax (VALAP) that is also used to attach the tube to the rotating wheel. The latter is driven by a phase-locked loop motor that can spin the wheel up to 100 Hz. The glass tube is placed at a distance of $R = 40$ mm from the center of the wheel. 

We use typical centrifugation frequencies of about $\nu=10$ Hz (600 rpm) corresponding to a centrifugal acceleration $a= (2 \pi \nu)^2 R \simeq 12 \times g$ (where $g \simeq 9.8 \, \mathrm{m/s^2}$ is the acceleration due to gravity). 
An estimate of the drift speed $v_d$ induced by centrifugal acceleration is given by $v_d = \mu \, f$, where 
$\mu$ is the mobility of the cell and $f$ is the resulting centrifugal force. 
For $a \simeq 12 \times g$ we can estimate $v_d \simeq 1 \, \mathrm{\mu m/s}$~\cite{mobility}.
This $v_d$ is significant when compared with the typical swimming speeds of \textit{E. coli} that are of order of $\sim 10\ \mathrm{\mu m /s}$~\cite{Berg,DDM3,DDM5}.


\section{Results and Analysis} \label{sec:res} 

\begin{figure}
\begin{center}
\includegraphics[width=8.5cm]{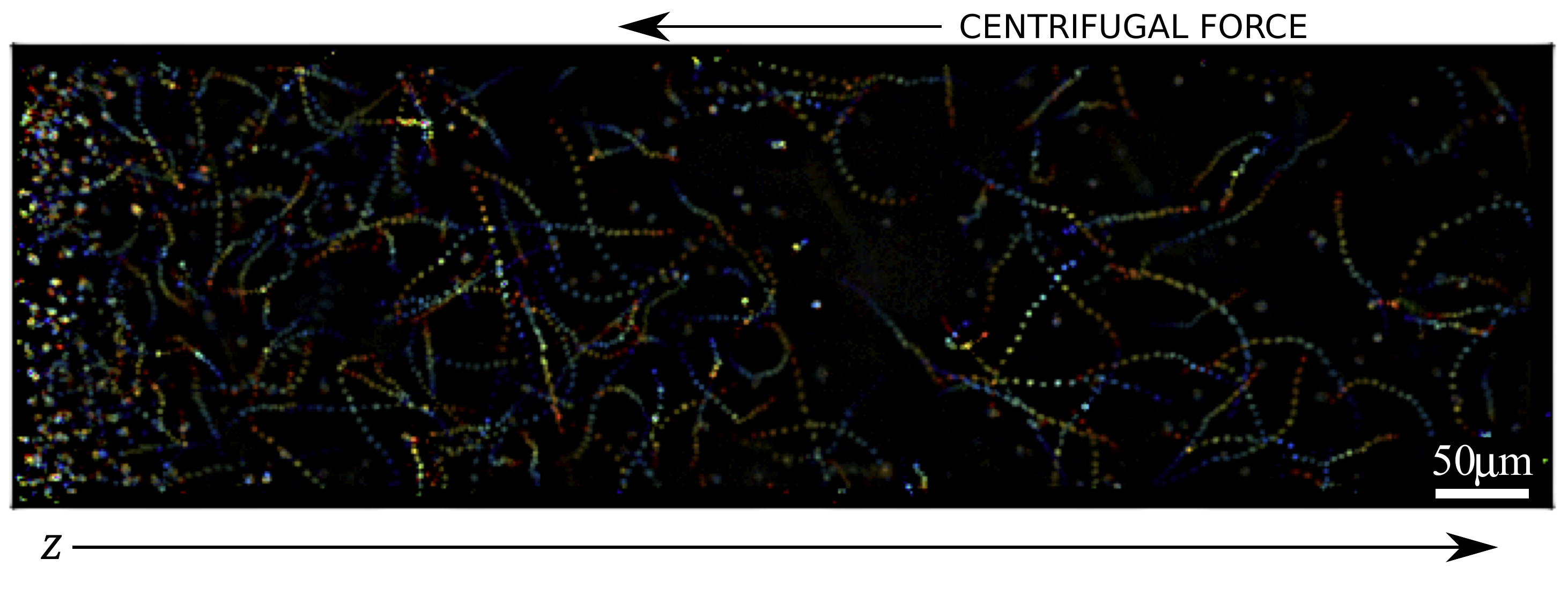}
\caption{ Traces of swimming bacteria after centrifugation at low cell density.
Slow bacteria are found sedimented at the bottom of the sample (left-hand side) while faster ones represent appear as the only visible component in the high-$z$ region of the sample.
{Traces are obtained as a superposition of frames that have been colored progressively from red to blue as time increases. }}
\label{fig:f1}
\end{center}
\end{figure}

We start by a qualitative study of bacterial trajectories at low cell concentration. The images are acquired after centrifugating for 10 minutes at $a \simeq 12 \times g$ and represent a total time lapse of 6 seconds. The centrifugal force field is parallel to the $z$ axis as indicated in Fig.~\ref{fig:f1}. 
{We choose to focus on cells swimming close to the capillary interface where most of the trajectories are confined by the wall and lay within the depth of field.}
As shown in Fig.~\ref{fig:f1} the non motile component is fully sedimented in a  $\sim 50 \mu$m thick pellet. Motile bacteria having a slow swimming speed appear as short trajectories mostly confined in the low $z$ side of the sample (left-hand side in Fig.~\ref{fig:f1}). In contrast, fast bacteria tracing longer paths, are found throughout the whole sample and they are clearly seen in the right-half of Fig.~\ref{fig:f1}. Interestingly the trajectories of fast bacteria are only rarely interrupted by sudden changes of direction (i.e. ``tumbles''~\cite{Berg}) signalling an average tumbling time that is longer than 6 seconds.

To quantify these effects we perform ICS on more concentrated samples ($10^9$ cells/ml).
ICS, as well as DDM, allows the use of digital video-microscopy to statistically characterize the dynamics of colloidal and/or biological samples~\cite{ICS1,DDM1,DDM2,DDM3,DDM4,DDM5} when images contain thousands of particles/cells, without tracking the motion of each individual element. 
To perform ICS we compute the image correlation function: 

\begin{equation} \label{eq:gqt}
g(\mathbf{q},t,t') = \langle I^\ast(\mathbf{q},t') \, I(\mathbf{q},t'+t) \rangle
\end{equation}

\noindent where $I(\mathbf{q},t)$ is the spatial Fourier transform of the image $I(\mathbf{r},t)$ at the wave-vector $\mathbf{q}$ 
and the star indicates the complex conjugate. 
Image acquisition is performed \textit{after} the centrifuge is stopped, i.e. while the swimming bacteria gradually start re-populating the whole sample. 
However, we can assume that, within the measurement time, the system is in a quasi-stationary state provided this re-diffusion process is much slower than the typical 
relaxation-time. In this approximation we can assume $g(\mathbf{q},t,t') = g(\mathbf{q},t)$ and therefore we can average over time-origins $t'$. The stationarity approximation has been checked by subdividing the 60s-long image acquisition in six different 10s sub-measurement. 
From these sub-measurement we can obtain six different $g(\mathbf{q},t)$ that do not show any systematic change with the measuring starting time.

{Centrifugation could, in principle, induce some anisotropy in the bacteria motion by aligning bacteria along the field direction.
However we take our measurements after several tenths of seconds from the stop of the centrifuge.
This ensures that bacteria have time to tumble and reorient substantially~\cite{isotropy}. This is also confirmed by looking at Fig.~\ref{fig:f1} where the trajectories appear randomized in all directions. In that situation we can also assume that the motion of bacteria is isotropic and $g(\mathbf{q},t)$ reduces to a function of $q=|\mathbf{q}|$ only, i.e.
$g(\mathbf{q},t)=g(q,t)$, so that we can average over all $\mathbf{q}$-vectors having the same magnitude. 
} {In the low density limit} the function $g(q,t)$  is connected to the self-intermediate scattering function $F(q,t)$ ~\cite{ICS1,DDM1,DDM2,DDM3}:

\begin{equation} \label{eq:gfqt}
g(q,t) = A(q)\, F(q,t)+B(q)
\end{equation}

{\noindent where the self-intermediate scattering function is defined as~\cite{Berne}: $F(q,t) = N^{-1} \sum_{j=1}^N \left\langle  \exp[i \, \mathbf{q} \cdot (\mathbf{r}_j(t)-\mathbf{r}_j(0))] \right\rangle$, $\mathbf{r}_j(t)$ being the position of the $j-$th cell at time $t$ and the sum is extended to all the $N$ cells in the image. $A(q)$ and $B(q)$ are time-independent factors depending on the average (static) structure of the image and on the background noise. {We have directly measured the relevant depth of field in our experiments by measuring the static structure factor $A(q)$ of cells that are stuck on the bottom interface while mechanically scanning the slide stage in the vertical direction. We found that, in the investigated $q$ range, only cells that lay within 40 $\mu$m from the focal plane contribute to the dynamic structure factor. Therefore, in order to avoid any contribution from bacteria swimming close to the bottom and top glass walls, from now on we will only discuss data that have been collected at the capillary middle plane.} {Note also that in DDM the term $B(q)$ becomes independent on static features in the images and it depends only on the background noise of the image, 
while in ICS this also contain information on the non-motile objects in the image. 
However most of the static features due to stuck cells are excluded from the depth of field when focusing away from the cell wall. Moreover centrifugation favours the elimination of such impurities. More importantly we study only on the time-varying part of the function $g(q,t)$ (that corresponds to the $F(q,t)$). We have checked that the $F(q,t)$ is practically identical when obtained by DDM or ICS.}

{
Following Ref.s~\cite{Berne,DDM3}, the $F(q,t)$ can be modelled as 
$F(q,t)=\mathrm{sinc}(q v t)$ if we assume that, on the relevant length-scales $\sim q^{-1}$, swimming bacteria move with speed $v$ along straight trajectories in 3D. In addition to that bacteria are subject to Brownian motion characterized by a diffusion constant $D$. This thermal motion alone would give $F(q,t)=\exp(-q^2 D t)$. Since the two processes are statistically independent they can be combined to get 
$F(q,t)=\exp(-q^2 D t) \ \mathrm{sinc}(q v t)$. Moreover some of the bacteria in the sample do not swim at all so that only a fraction $\alpha$ of motile cells is present. In this case the function becomes $F(q,t)=\exp(-q^2 D t) [\alpha \, \mathrm{sinc}(q v t)+(1-\alpha)]$.
Finally, considering the fact that the swimming speed $v$ may vary a lot among the bacterial population, the $F(q,t)$ can be rewritten as:
}

\begin{equation} \label{eq:fqt}
F(q,t) = e^{-q^2 D t} \left[ \alpha \int_0^\infty dv \, P(v) \, \mathrm{sinc}(q v t) +(1-\alpha) \right]
\end{equation}

\noindent where $P(v) \, dv$ represents the probability of 
finding a bacterium with speed between $v$ and $v+dv$. 

Rather than assuming a specific functional form for the $P(v)$ ~\cite{DDM3,DDM5}, 
we model $P(v)$ as a superposition of triangular functions defined on a grid of overlapping nodes.~\cite{Splines}. This can be seen as a spline-interpolating curve approximating the actual $P(v)$. 
The resulting $P(v)$ is continuous, positive 
and bounded by some maximum speed $v_\mathrm{max}$ (i.e. $P(v>v_\mathrm{max})=0$).
The spline $P(v)$ is thus written as

\begin{equation} \label{eq:spline}
P(v)=\mathcal{N}^{-1} \sum_{j=1}^M c_j \, T_j(v)
\end{equation} 

\noindent where the $T_j(v)$ are the triangular functions, each having amplitude $c_j$, 
and the sum is taken over the $M$ functions used and the normalization factor $\mathcal{N}=\sum_{j=1}^M c_j \, \int dv \, T_j(v)$ ensures that $\int dv P(v)=1$.
{We choose to set the first knot at $1$ $\mu$m/s so that the non-motile component is fully accounted by the $1-\alpha$ term in Eq.~(\ref{eq:fqt}). We used 13 knots
spaced by approximately 5 $\mu$m/s (similarly to Ref.~\cite{Splines} ).
We have checked that varying the number of knots from this value to about 30 does not affect significantly the obtained fitted distributions (see Supplemental Material).}
Putting together (\ref{eq:gfqt}),  (\ref{eq:fqt}) and ({\ref{eq:spline}) we can determine $P(v)$ by a fitting procedure having as parameters: $A(q)$, $B(q)$, $D$, $\alpha$  and the set of spline coefficients $c_j$. The fitting parameters $D$, $\alpha$ and $c_j$ do not depend on $q$, in principle, and in fact only display weak variations in the range $0.4 \leq q \leq 1.15 \, \mathrm{\mu m^{-1}}$. 
Therefore we choose to obtain $P(v)$ as an average over $q$ values in that range.
We expect that the presence of an acceleration field along $z$ will lead to a spatial inhomogeneuos  dynamics along the same axis. Therefore we divide each image in seven sub-images ($1200 \times 256$ pixels, i.e. $\sim 0.44 \times 0.093 \ \mathrm{m m}$) along the $z$ axis and separately compute a $g(q,t)$ function for each of them. As a result we obtain a set of seven intermediate scattering functions $F(q,t)$. 
We choose to focus on the high-$z$ half of the sample where we expect to find the 
fastest cells after centrifugation. In this configuration the highest $z$ and the lowest $z$ included in the image are $z=0.95 \, \mathrm{m m}$ and $z=0.39 \, \mathrm{m m}$ respectively.

\begin{figure}
\begin{center}
\includegraphics[width=9.5cm]{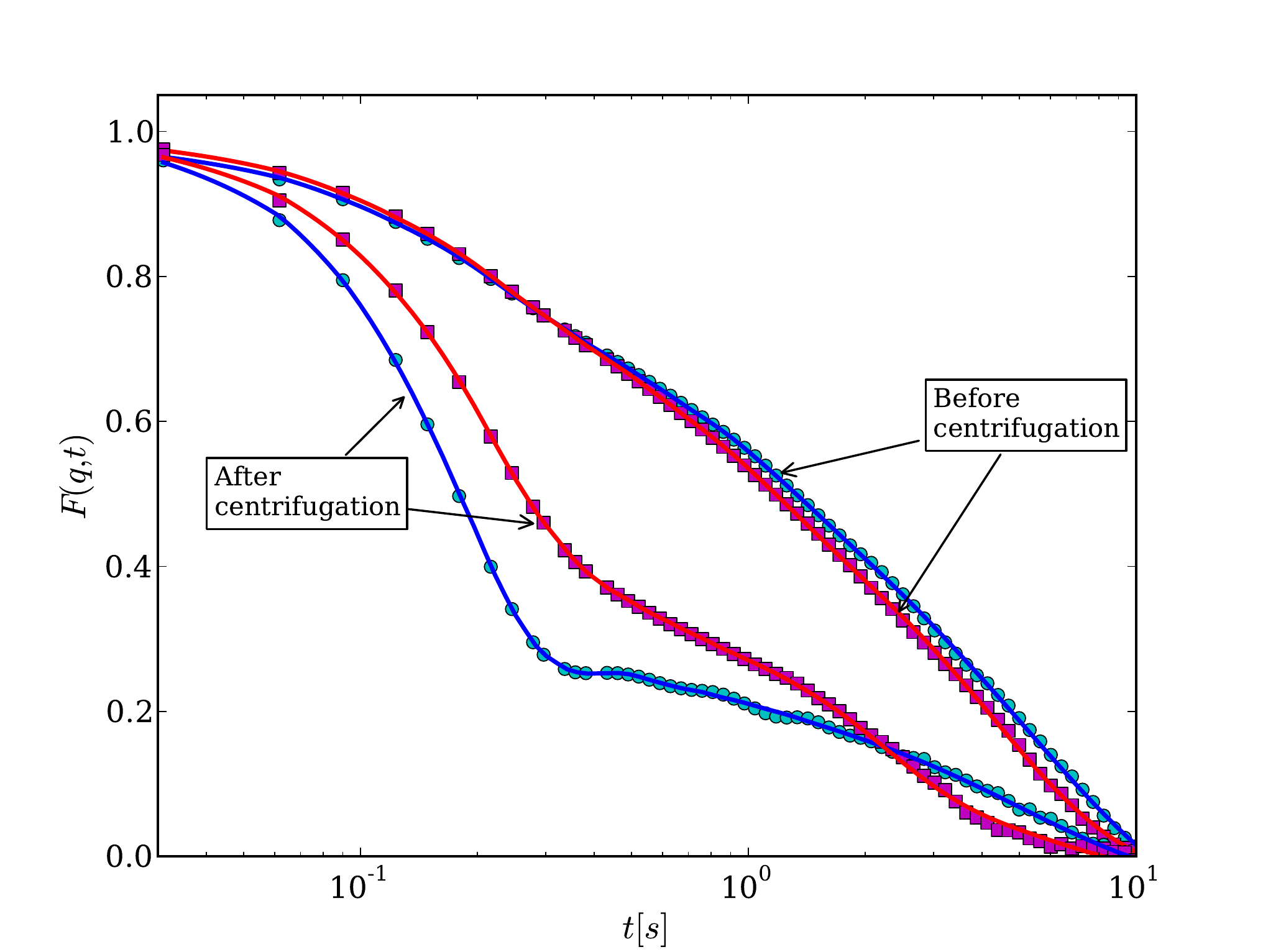}
\caption{Self-intermediate scattering functions ($q=0.77 \, \mathrm{\mu m^{-1}}$) 
obtained from ICS, before and after centrifugation, at $z=0.39 \, \mathrm{m m}$ ($\Box$) and 
$z=0.95 \mathrm{m m}$ ({\Large $\circ$}).
The two $F(q,t)$ relaxing faster are measured after a 10 min-long centrifugation at $a \simeq 12 g$. 
The full lines are fits with the model function of Eq.~(\ref{eq:fqt}).
Note that in the $a=12 \, g$ case the $F(q,t)$ relaxes faster at $z=0.95 \, \mathrm{m m}$ than at $z=0.39 \, \mathrm{m m}$ indicating that the average speed of bacteria is higher at larger $z$.
}
\label{fig:f2}
\end{center}
\end{figure}
  
In Fig.~\ref{fig:f2} we show the $F(q,t)$, for one single $q=0.77 \, \mathrm{\mu m^{-1}}$, at two different $z$ before and after centrifugation. 
The fitted functions are represented by full lines.
Fig.~\ref{fig:f2} shows quite clearly how the functions measured after a 10 min-long centrifugation at $a \simeq 12 g$ relax faster than the functions measured in the unperturbed sample ($a=0$). The two $F(q,t)$, measured at two different $z$, in the unperturbed sample are very similar. Differently, in the centrifuged case we clearly see that the $F(q,t)$ measured at high $z$ relaxes faster than the $F(q,t)$ measured at low $z$ suggesting that average speed of the bacteria is higher at larger $z$. In addition, after centrifugation, the amplitude of the motile component increases from $\alpha=0.2$ to $\alpha=0.7$ at  higher $z$ values. Throughout the range of centrifugations speeds and $z$ values we always find diffusivities of $D=0.3\pm0.1 \, \mathrm{\mu m^2/s}$ that are consistent with the expected values for Brownian motion of non motile cells \cite{DDM5}.

\begin{figure}
\begin{center}
\includegraphics[width=9.5cm]{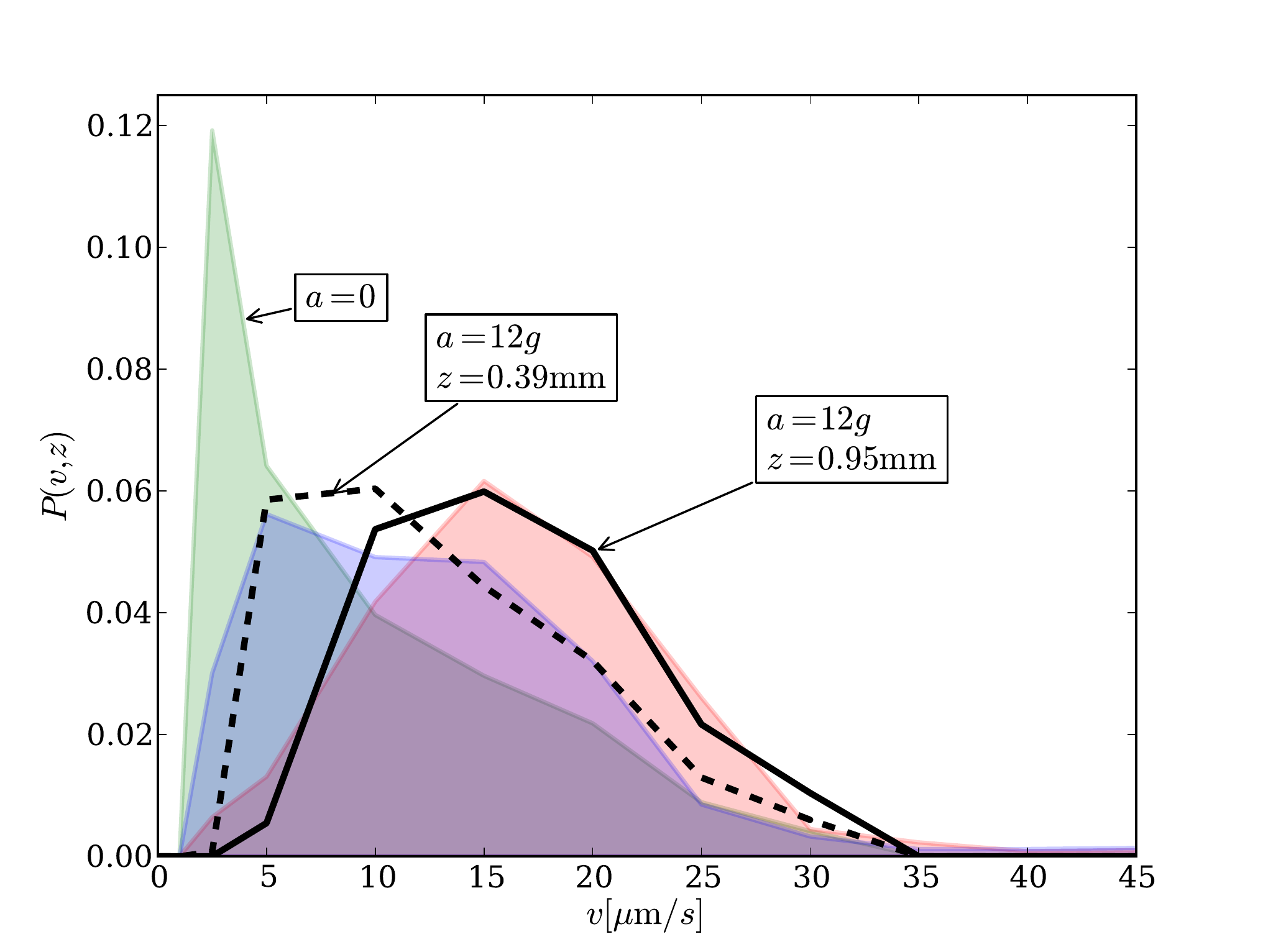}
\caption{ 
Speed probability density functions measured in the uncentrifuged ($a=0$) and the centrifugated sample ($a=12 g$) 
represented by the shaded areas. 
The $P(v,z)$ of the centrifuged sample is displayed at two $z$, $z=0.95 \, \mathrm{m m}$ and $z=0.39 \, \mathrm{m m}$ (as indicated by the arrows) obtained from the fits of the two $F(q,t)$ of Fig.~\ref{fig:f2}. Note that the $P(v,z)$ shift to higher speed and changes  shape at higher $z$. The dashed and the full line are, respectively, the $P(v,z)$ at $z=0.36 \, \mathrm{m m}$ and at $z=0.95 \mathrm{m m}$, obtained from the $P_0(v)$ of the uncentrifuged sample, using the model presented in Sec.~\ref{sec:model}.
}
\label{fig:f3}
\end{center}
\end{figure}

In Fig.~\ref{fig:f3} we plot the speed distributions before and after centrifugation. The original speed distribution of the unperturbed  sample is denoted by $P_0(v)$. After centrifugation speed distributions acquire a dependence on $z$ which is now explicitly introduced as a parameter in $P(v,z)$. In Fig. \ref{fig:f3} we show two speed distributions $P(v,z)$ corresponding to the same two $z$ values in Fig.~\ref{fig:f2}.
Note that the initial $P_0(v)$ shows a very high concentration of slow bacteria.
Differently the $P(v,z)$ at $a=12 g$ is peaked around higher speeds demonstrating that centrifugation has sedimented the most of the slow bacteria at low $z$. 
Note also that the $P(v,z)$ at $z=0.95 \, \mathrm{m m}$ is shifted to higher speeds with respect to the $P(v,z)$ at $z=0.39 \, \mathrm{m m}$. Interestingly the $P(v,z)$ at $z=0.39 \, \mathrm{m m}$ is quite similar to a Shultz distribution~\cite{DDM3,DDM5}, while at $z=0.95 \, \mathrm{m m}$ we find a $P(v,z)$ displaying a more symmetric shape.

\begin{figure}
\begin{center}
\includegraphics[width=9.5cm]{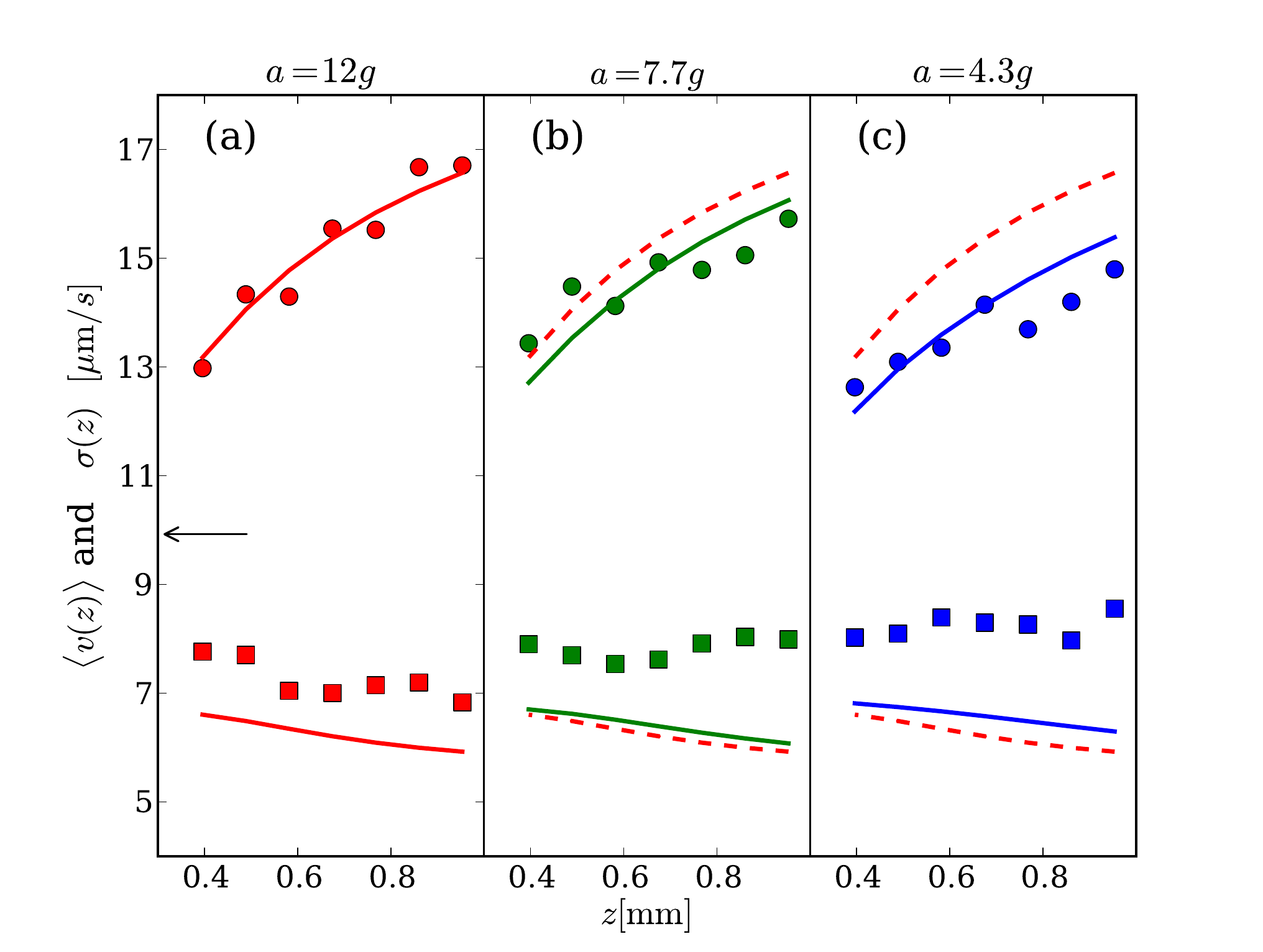}
\caption{ 
(\textbf{a}) Average speed $\langle v(z) \rangle$ ({\Large $\circ$}) and speed standard deviation $\sigma(z)$ ($\Box$) as function of $z$ after centrifugation at $a \simeq 12 \, g$, corresponding 
to the same measurement of Fig.~\ref{fig:f2},\ref{fig:f3}. 
The arrow represents the mean speed (full line) of the uncentrifuged sample. 
Note how $\langle v(z) \rangle$ increases upon increasing $z$.
The full curve is obtained by a heterogeneous diffusion model, 
discussed in Sec.~\ref{sec:model}.
(\textbf{b}) and (\textbf{c}) $\langle v(z)\rangle$ ({\Large $\circ$}) and $\sigma(z)$ ($\Box$) as a function of $z$ after centrifugation at $a \simeq 7.7 \, g$ and $4.3 \, g$ respectively. The full curve in (a) is reproduced in (b) and (c) as a dashed curve for an easier comparison.
This shows how the $\langle v(z)\rangle$ shifts to lower speeds upon decreasing $a$.
}
\label{fig:f4}
\end{center}
\end{figure}

From the $P(v,z)$ we can determine the average speed $\langle v(z) \rangle=\int dv \, P(v,z) \, v$. The $\langle v(z) \rangle$, corresponding to the measurement of Fig.~\ref{fig:f1} and \ref{fig:f2}, is shown in Fig.~\ref{fig:f4} (a) for all values of $z$ between $0.36$ and $0.95 \, \mathrm{m m}$. $\langle v(z) \rangle$ grows from about $13 \, \mathrm{\mu m/s}$ to almost $17 \, \mathrm{\mu m/s}$ as $z$ increases. 

In Fig.~\ref{fig:f4} (b) and (c) we also show the effect of reducing the centrifugal acceleration $a$ applying centrifugation to two additional samples prepared from the same (not centrifuged) original sample. The $\langle v(z) \rangle$ in Fig.~\ref{fig:f4}(b) and (c) were measured after 10 min-long centrifugation with $a\simeq7.7 \, g $ and $4.3 \, g$ respectively. It can be seen that the whole $\langle v(z) \rangle$ profile shifts to lower speeds upon decreasing $a$. 

In addition to the average speed we can study the behaviour of the standard deviation of the speed ${\sigma(z)}$ that is given by ${\sigma(z)}^2=\int dv \, P(v,z) \, (v-\langle v(z)\rangle)^2$. The $\sigma(z)$, corresponding to the measurement of Fig.~\ref{fig:f1} and \ref{fig:f2}, is shown in Fig.~\ref{fig:f4} (a) as a function of $z$. It can be seen that $\sigma(z)$ has a much weaker dependence on $z$ with respect to $\langle v(z) \rangle$. $\sigma(z)$ decreases only of about $1 \mathrm{\mu m/s}$ upon increasing $z$.
This is also evident from  Fig.~\ref{fig:f3} where we can see that the $P(v,z)$ shifts to higher speeds and changes shape as $z$ increases but its width remains quite large. Moreover $\sigma(z)$ weakly increases while decreasing the centrifugal acceleration as shown in Fig.~\ref{fig:f4} (b) and (c).

{ In the case of {\it E. coli} bacteria careful procedures have been developed to start off with a suspension of cells having high motility characteristics. Typical speed distributions~\cite{DDM5} compare well with what we observe in the small $z$ fraction of the centrifuged sample with a relative width ($\sigma / \langle v \rangle$) of 0.6. The high $z$ component of the centrifuged sample further improves on that motility leading to a distribution having a higher average and a smaller relative width 0.4. Moreover, as discussed in the following section, our method works, in principle, for any kind of self propelled object and could be particularly useful in the case of biological or chemical swimmers that can only be produced with highly polydisperse speeds.
}

\section{Modelling} \label{sec:model} 

The behaviour of swimming cells under the action of an external field was recently subject of accurate theoretical investigation~\cite{CatesPRL,CatesSed,Lorentz}. 
Building on experimental observations~\cite{Berg}, in these works, the motion of the swimming bacterium was modelled as a sequence of straight ``runs'' at constant speed $v$ interrupted sudden reorientations, named ``tumbles'', occurring randomly with rate $\lambda$. This dynamics leads to a diffusive behaviour at long times in absence of external perturbations. 

When a ``run and tumble'' cell is subject to an external force field this changes the probability of finding the cell in a given position in space. 
In the case of interest here a centrifugal acceleration $a$, directed along the $z$-axis, adds a drift-speed component $v_d\hat{\mathbf{z}}$ to the velocity $\mathbf{v}$ of the cell. If this external field is applied for a very long time the swimming cells will assume a stationary density profile. In this situation the probability $\rho(z, v)dz$ of finding a bacterium with speed $v$ in the range $z, z+dz$, has an exponential form~\cite{CatesSed}: 

\begin{equation} \label{eq:rho1}
\rho(z, v) = \mathcal{N}^{-1} \exp(- \kappa \, z)
\end{equation} 

\noindent where the dependence on $v$ is implicit in the sedimentation rate $\kappa$ and $\mathcal{N}=\int_0^H \exp(- \kappa \, z)$ is a normalization constant obtained by integrating over the height of the sample $H$. The actual value of $\kappa$ depends on the speed of the bacterium, its tumbling rate and the drift speed according to the transcendental equation:

\begin{equation} \label{eq:kappa1}
\ln \left[ \frac{\kappa (v+v_d)+\lambda}{\kappa (v-v_d)+\lambda} \right] = \frac{2\kappa v}{\lambda}
\end{equation}

\noindent It can be shown that, in the limit of small $v_d$,  $ \kappa = v_d/D_\mathrm{eff}$ where $D_\mathrm{eff}=v^2/(3\lambda)$ is the effective diffusion coefficient of run and tumble dynamics. In this case, Eq.(\ref{eq:rho1}) reduces to the Boltzmann form in Eq.(\ref{boltz}) (with $D=D_\mathrm{eff}$):

\begin{equation} \label{eq:rho2}
\rho(z,v) = \mathcal{N}^{-1} \exp \left(- \frac{v_d}{D_\mathrm{eff}} \, z \right)=\mathcal{N}^{-1} \exp \left(-3\frac{ v_d \lambda}{v^2}z \right)
\end{equation} 

We have checked numerically that the approximated $\rho(z)$ of Eq.(\ref{eq:rho2}) gives practically the same results of Eq.(\ref{eq:rho1}) in the range $z$ of interest here and with the drift speeds ($\sim 1 \mathrm{\mu m /s}$) applied in the experiment.

To quantitatively account for the distribution of speeds in the bacteria population we consider the probability density function of the speed $P_0(v)$ in absence of any external field. 
As mentioned in the previous Section this function is measured experimentally in the uncentrifuged sample.
If the centrifugal field is applied the probability of finding a bacterium with speed within $(v,v+dv)$ and located between $z$ and $z+dz$ is given by $P_0(v) \, \rho(z,v) \, dv \, dz$. From this expression we see that the function 
$\rho(z,v)$ acts as a filter that reduces the probability of finding slow bacteria in the high-$z$ regions. In this framework the $z$-dependent speed distribution is given by

\begin{equation} \label{eq:pvT}
P(v,z) = \mathcal{N}^{-1} P_0(v) \, \rho(z,v)
\end{equation}

\noindent where $\mathcal{N}=\int dv \, P_0(v) \, \rho(z,v)$. 
Since $v_d$ and $\lambda$ appear as product in Eq.(\ref{eq:rho2}) we choose $\gamma = \lambda v_d$ as the only free fitting parameter.
To find $\gamma$ we fit the average speeds at different $z$ measured after centrifugation at $a=12g$ with the function $\langle v(z)\rangle=\int \mathcal{N}^{-1} P_0(v) \, \rho(z,v) $.
This fit is represented by the full curve in Fig.~\ref{fig:f4}(a) giving the best fit parameter 
$\gamma = 4.5\times 10^{-2} \mathrm{\mu m /s^2}$.
This value of $\gamma$ is compatible for example with the values $v_d=0.45 \, \mathrm{\mu m/s}$ and $\lambda = 0.1$ Hz. This $v_d$ is of the same order of magnitude of the estimate given above~\cite{mobility} and this tumbling rate $\lambda$ is consistent with the direct imaging of bacteria trajectories, discussed at the beginning of  Sec.~\ref{sec:res}. We recall indeed that the traces of bacteria displayed in Fig.~\ref{fig:f1} show that cells swim along roughly straight runs for about 6~s without evident signs of tumbling. The observed value for the tumbling rate is about an order of magnitude smaller than what generally reported in literature~\cite{Berg}. It will be important to investigate the presence of an heterogeneous tumbling frequency among our cells and the  possible role of centrifugation in selecting the smoothest swimmers.

In Fig.~\ref{fig:f3} we show the 
$P(v,z)$ at $z=0.95 \, \mathrm{m m}$ (full curve) and the $P(v,z)$ at $z=0.36 \, \mathrm{m m}$ (dashed curve) obtained from Eq.s (\ref{eq:rho2}) and (\ref{eq:pvT}) using the fitted parameter $\gamma = 4.5 \times 10^{-2} \mathrm{\mu m /s^2}$.
Note that the modelled $P(v,z)$ follow quite well the measured function at that $z$. 
When computing $\sigma(z)$ we also find a slightly decreasing trend with increasing $z$ although experimental values are systematically higher (see Fig.~\ref{fig:f4}(a)). 
The model predicts that the $\sigma(z)$ should decrease upon increasing $z$ and the data seem to follow this trend.

To further test the model we have computed the quantities of interest at different centrifugal accelerations. 
Since the drift speed, and hence $\gamma$, are proportional to the centrifugal acceleration ($v_d \propto a$)  we can obtain the $P(v,z)$ at other values of $a$ by simply rescaling $\gamma$.
The $\langle v(z)\rangle$ obtained in this way are represented by the full curves in Fig.~\ref{fig:f4} (b) and (c) for the centrifugal accelerations $a=7.7 \, g$ and $4.3 \, g$ respectively. 
{ The model has now no fitting parameters and shows an overall decreases of $\langle v(z)\rangle$ with decreasing $a$ that is in quantitative agreement with the data shown Fig.~\ref{fig:f4} (b) and (c). When the parameter $\gamma$ is set free in the fitting this is found to be approximatively the same at all centrifugal accelerations ($\gamma \approx 4-5 \times 10^{-2} \mathrm{\mu m /s^2}$).}
Within the same calculation we can estimate the $\sigma(z)$ at $a=7.7 \, g$ and $4.3 \, g$. The theoretical values are represented by the full curves in the lower part of Fig.~\ref{fig:f4} (b) and (c) and they are still systematically smaller than the measured ones. This could be due to an intrinsic fluctuation of swimming speeds of individual cells.

\section{Conclusions} 

Summarizing our findings, we have demonstrated that centrifugation induces significant effects on the spatial distribution of motility characteristics of \textit{E. coli} cells. 
By using dynamic image correlation spectroscopy we have found that a substantial speed gradient is generated along the direction of the centrifugal acceleration. Finally we have compared our results with a theoretical model describing bacteria as particles with heterogeneous diffusion constant. 
When compared to other strategies for motility sorting, such as those based on microfabrication, centrifugation offers some important advantages like:  easy tunability, large sample volumes, fast operation and easy implementation. 

Our findings could be further used to design novel microfluidic procedures aiming at sorting out highly motile cells from heterogeneous populations. 
This would be important both for micro engineering applications that exploit self-propulsion or for further investigation of the biological factors associated to a high motility.

The research leading to these results has received funding from the European Research Council under the European Union's Seventh Framework Programme (FP7/2007-2013) / ERC grant agreement n$^\circ$ 307940. We also acknowledge funding from IIT-SEED BACTMOBIL project and MIUR-FIRB project
No. RBFR08WDBE.


\begin{thebibliography}{99}
\bibitem{mobility} Following Ref.~\cite{Image} we approximate the cell body with a prolate ellipsoid and the
flagellar bundle~\cite{Bundles} as an helix obtaining a mobility $\mu \simeq 30 \, \mathrm{\mu m/(s \, pN)}$. The centrifugal force $f$ acting on one single bacterium is $f=\Delta m \, a$ where $a$ the centrifugal acceleration and $\Delta m$ is the buoyant mass of the cell that can be expressed as $\Delta m= \Delta \rho \, V $, $\Delta \rho$ being the excess density of the cell with respect to the density of the medium, and $V$ being the volume of the cell. By taking the $\Delta \rho$ measured in Ref.~\cite{ColiDens} we can estimate $v_d \simeq 1 \, \mathrm{\mu m/s}$ at $\nu = 10$ Hz.
\bibitem{isotropy} {
Note that the measurement are taken after about one minute from the stop of the centrifuge.
If we assume that the reorientation of the bacteria is a random event happening at constant rate $\lambda$
the fraction of bacteria that have not tumbled from at the beginning of the measurement is only few percents 
for $\lambda$ as low as $0.05$ Hz. This can be easily estimated from the Poisson distribution.
}
\end{thebibliography}
\end{document}